\documentclass[aps,amsmath,amssymb,nofootinbib,twocolumn,a4paper,prb,,showpacs]{revtex4-1}
\usepackage{verbatim}
\usepackage{graphicx}
\usepackage{color}
\usepackage{multirow}
\usepackage{subfig}
\usepackage{float}
\usepackage{caption}
\newcommand{\note}[1]{$\color{red}{\bullet}$}
\begin{document}

\title{\textbf{Spinful Composite Fermions in a Negative Effective Field}}
\author{Simon C. Davenport and Steven H. Simon}
\affiliation{Rudolf Peierls Centre for Theoretical Physics, 1 Keble Road, Oxford, OX1 3NP, United Kingdom}
\date{\today}

\begin{abstract}

In this paper, we study fractional quantum Hall composite fermion wavefunctions at filling fractions $\nu = 2/3 , 3/5,$ and $4/7$. At each of these filling fractions, there are several possible wavefunctions with different spin polarizations, depending on how many spin-up or spin-down composite fermion Landau levels are occupied. We calculate the energy of the possible composite fermion wavefunctions and we predict transitions between ground-states of different spin polarizations as the ratio of Zeeman energy to Coulomb energy is varied. Previously, several experiments have observed such transitions between states of differing spin polarization and we make direct comparison of our predictions to these experiments. For more detailed comparison between theory and experiment, we also include finite-thickness effects in our calculations. We find reasonable qualitative agreement between the experiments and composite fermion theory.  Finally, we consider composite fermion states at filling factors $\nu = 2+2/3 , 2+3/5$, and $2+4/7$. The latter two cases we predict to be spin polarized even at zero Zeeman energy.

\end{abstract}


\maketitle

\section{Introduction}

The distinctive band structure of two-dimensional electrons in a magnetic field has proved to be a rich setting for new and exciting physical phenomena. In particular, in the absence of interactions or disorder, the spectrum of single particle eigenstates breaks into degenerate bands called Landau levels (LLs); fractional quantum Hall (FQH) physics\cite{prangebook} occurs when interactions between electrons break the degeneracy of partially filled Landau bands, leading to an incompressible fluid ground-state. Quantum Hall systems are typically characterized by a filling factor $\nu$, which quantifies the ratio of the number of electrons in the two-dimensional electron gas (2DEG) to the number of flux quanta through the sample or, alternatively, the filling factor quantifies the number of filled LLs in the system.

Naively, one might expect that the high magnetic field characteristic of the fractional quantum Hall regime might remove any spin degree of freedom entirely. However  in conventional GaAs systems, owing to the small $g$ factor, even in fairly high fields, unpolarized or partially polarized quantum Hall states may occur.    The favorability of spin non-polarized states is generally determined by the ratio of the Zeeman energy per electron, $E_Z = g \mu_B B$ the energy associated with flipping a spin in magnetic field strength $B$, to the Coulomb energy per electron $E_{C}$, the energy associated with the spatial configuration of electrons in the quantum Hall system.   For many quantum Hall states (particularly composite fermion states) due to the Pauli exclusion principle, the Coulomb energy (or, equivalently, the effective composite fermion kinetic energy) can be lower if spins are not fully polarized.  As a result, if the Zeeman energy is not too large, quantum Hall ground-states may not be fully spin polarized.

In a number of recent  experiments the degree of spin polarization of quantum Hall fluid has been explicitly measured as a function of the ratio of Zeeman to Coulomb energy for a variety of fixed filling factors. \cite{groshaus2007,kukushkin1999,khandelwal1998,cho1998,manfra1996,du1995}  In this paper, we will focus on Refs.~\onlinecite{kukushkin1999} and \onlinecite{du1995}, where detailed comparison to our theoretical work is possible.

In Ref. \onlinecite{kukushkin1999} a non-trivial net spin polarization is observed at the following filling factors in the lowest Landau level (LLL): $\nu = 2/3 , 3/5, 4/7 ,2/5 , 3/7$ and $4/9$. In these experiments the aim is to keep the filling factor fixed while varying the ratio of the Zeeman to the Coulomb energy.  This is achieved by varying both the applied field strength and the density of electrons in the sample, keeping their ratio (hence $\nu$) fixed. For each filling factor, as the field strength is varied, the experiments report a set of plateau with constant spin polarization, punctuated by a series of relatively sharp transitions. Similar conclusions can be inferred from data presented in Ref.~\onlinecite{du1995} at filling factors $\nu = 4/3, 7/5, 10/7, 8/5, 11/7$ and $14/9$. For high field and high electron density, the spin is polarized, but as the field strength and electron density are reduced, there are transitions to FQH states of successively smaller net spin polarization until, finally, some lower bound of spin polarization is reached, with the particular lower bound depending on the filling factor. These transitions are observed to occur at some critical values of the applied field $B^{\mbox{\tiny crit}}$ and there is, therefore, a corresponding critical Zeeman energy per electron, $E^{\mbox{\tiny crit}}_Z = g \mu_B B^{\mbox{\tiny crit}}$.

From a theoretical perspective, an exceptional phenomenological understanding of the fractional quantum Hall effect (FQHE) has been acquired via the concept of the composite fermion (CF). \cite{jainbook} The key notion of CF theory is that the problem of strongly interacting electrons in a perpendicular magnetic field can be mapped onto the problem of non-interacting CFs in an effective magnetic field. The direction of the effective magnetic field can be either parallel or antiparallel to the physical magnetic field. The CFs can be pictured as electrons bound to a certain number, $p$, of magnetic flux quanta.  In the effective magnetic field, there exist effective Landau levels---these are completely analogous to the LLs that occur for noninteracting electrons in the presence of a magnetic field. The FQHE of electrons can be interpreted as the integer quantum Hall effect of CFs with $p$ fluxes attached and occupying a certain number $n$ of the effective LLs; this idea has proved to be very successful (see, for example, J. K. Jain's book, Ref.~\onlinecite{jainbook}). The principal set of filling factors encompassed by CF theory are given by
\begin{equation}
\label{fillingfactor}
\nu = n/(2pn \pm 1)
\end{equation}
where the sign here indicates the direction of the effective field relative to the real magnetic field. Taking into account the two spin species, particle-hole conjugate versions of these states occur at filling factors $2-\nu$.

In CF theory, trial quantum Hall states with a spin degree of freedom can be constructed by simply associating a spin degree of freedom with the CFs themselves. \cite{jainbook,park2001} CFs of each spin species can independently occupy a non-negative integer number  $n_\uparrow$ and $n_\downarrow$ of effective LLs; the filling factor remains as in Eq.~(\ref{fillingfactor}), but now with $n=n_\uparrow +n_\downarrow$.  Consequently a whole series of CF wavefunctions is possible at each filling factor (a visualization of such a series of states can be found in Refs. \onlinecite{jainbook} or \onlinecite{park2001} and a modified version is presented here in Fig. \ref{qualitativepredictions}).   
 
The experiments of Ref.~\onlinecite{kukushkin1999}  examined the $p=1$ series for $n=2,3, 4$ with both positive and negative effective magnetic fields:  $\nu=2/5 , 3/7$ and $4/9$ in the positive effective magnetic field case and $\nu = 2/3 , 3/5$ and $4/7$ for the negative effective field case. Ref.~\onlinecite{du1995} examined the particle-hole conjugates of these states at filling factors $2-\nu$.  In each case, qualitatively at least, the predictions of CF theory appear to support experimental observations. We shall elaborate more on these qualitative predictions in Sec. \ref{wavefunctions}.

Transitions between CF ground-states with different spin polarizations can occur when the difference in Coulomb energy per electron between the two ground-state CF configurations compensates for the increase in Zeeman energy per electron due to the spin depolarization.  The differences in the Coulomb energies per electron of various spin CF states at the same filling factor thus can be related to the critical Zeeman energy per electron for transitions between the various spin-polarizations of the quantum Hall fluid. Calculating the Coulomb energy per electron for composite fermion trial wavefunctions, Park and Jain\cite{park2001} were then able to directly compare the experimental measurements and theoretical predictions of the critical Zeeman energies per electron for filling factors $\nu=2/5 , 3/7$ and $4/9$ (the positive effective field case). Their results for the predicted values of the critical Zeeman energy per electron for these three filling factors, for the most part, agree well with the values measured in both Refs.~\onlinecite{du1995} and \onlinecite{kukushkin1999}. These authors did not perform the same calculation and comparison for filling factors $\nu = 2/3 , 3/5$ and $4/7$ (the negative effective field case). In this work we will present the results of our calculations for the negative effective field case, we shall evaluate the critical Zeeman energies for transitions between different spin states and then we will compare our results to the relevant experimental measurements.

CFs have been the subject of a great deal of investigation and there are now well-developed techniques which have been established for calculating the associated Coulomb energy using the well-known Metropolis Monte Carlo procedure (see, e.g., Refs. \onlinecite{jain1997,park2001, moller2005}). Nevertheless, for the series of states of interest here, with the exception of the spin polarized cases (see Ref.~\onlinecite{moller2005}), the Coulomb energies have not yet been calculated. For sufficiently large numbers of particles the numerical evaluation of the trial wavefunctions with negative effective field turns out to be highly non-trivial---and much more complex than the positive effective field case. We have constructed an efficient new numerical algorithm to handle this situation. The details of the algorithm are discussed in Appendix \ref{CFalgorithm}.

In this work, we calculate the interaction energy associated with the CF trial wavefunctions using a simple Coulomb interaction potential, what we have called the Coulomb energy.  Such an interaction would apply to a perfectly 2D geometry; however, a laboratory quantum Hall system cannot be considered perfectly 2D and more realistic model interactions must take into account the finite extent of the system in the direction perpendicular to the 2D plane. To study the effect of such a modification to the theory, we implement an interaction which takes into account finite-thickness effects (see, e.g., Ref.~\onlinecite{peterson2008}). Finally, we calculate the Coulomb energy appropriate for the 2nd LL analogies of the LLL CF trial wavefunctions discussed previously (the 2nd LL being the next LL above the LLL i.e., filling factors $\nu=\nu_{\mbox{ \tiny LLL}}+2$). Using these results we make an additional prediction that for CF trial ground-states it is not energetically favorable to have non-polarized states in the 2nd LL with negative effective magnetic field.

The structure of this paper is as follows:  in Sec. \ref{wavefunctions} we shall briefly summarize the qualitative predictions of CF theory and then we shall write down the explicit forms of the CF wavefunctions which we are interested in. In Sec. \ref{results} we shall present the results for the Coulomb energy of the various ground-state trial wavefunctions in the LLL and 2nd LL, the critical Zeeman energy predicted by CF theory for the LLL and 2nd LL and the results of the finite thickness correction. We shall also compare our results to the experimentally measured values of the critical Zeeman energy in the LLL.  Finally, we shall make some remarks on our findings in Sec. \ref{conclusion}.



\section{Theory of Composite Fermions with Spin}
\label{wavefunctions}

\subsection{Qualitative predictions of CF theory}

Before we describe the explicit forms of the CF trial wavefunctions,  we shall first briefly summarize the qualitative predictions made by CF theory. The filling factors of interest correspond to CFs with $p=1$ flux attached and occupying $n=$~2--4 effective LLs. In the case of negative effective field, using Eq.~(\ref{fillingfactor}), these are $\nu = 2/3 , 3/5, 4/7$, respectively, and in the case of positive effective field these are $\nu=2/5, 3/7$, and $4/9$, respectively. For each filling factor the set of possible spin-dependent states is deduced by considering all possible non-negative integer values of  $n_{\uparrow}$ and $n_{\downarrow}$ satisfying $n=n_{\uparrow}+n_{\downarrow}$. For a system of $N_{\uparrow}$ spin-up electrons and $N_{\downarrow}$ spin-down electrons, we shall define the ``degree of spin polarization'' by $\gamma_e = \frac{N_{\uparrow}-N_{\downarrow}}{N_{\uparrow}+N_{\downarrow}} $ and in the thermodynamic limit, where each effective LL contains the same number of electrons, it can be shown that  (see Appendix \ref{extrastuff})
\begin{equation}
\label{degreeofspin}
\gamma_e =\frac{n_{\uparrow}-n_\downarrow}{n_\uparrow+n_{\downarrow}}.
\end{equation}

\begin{widetext}

\begin{figure}[t]
\includegraphics[trim = 0mm 40mm 0mm 20mm, clip, width=0.5\textwidth]{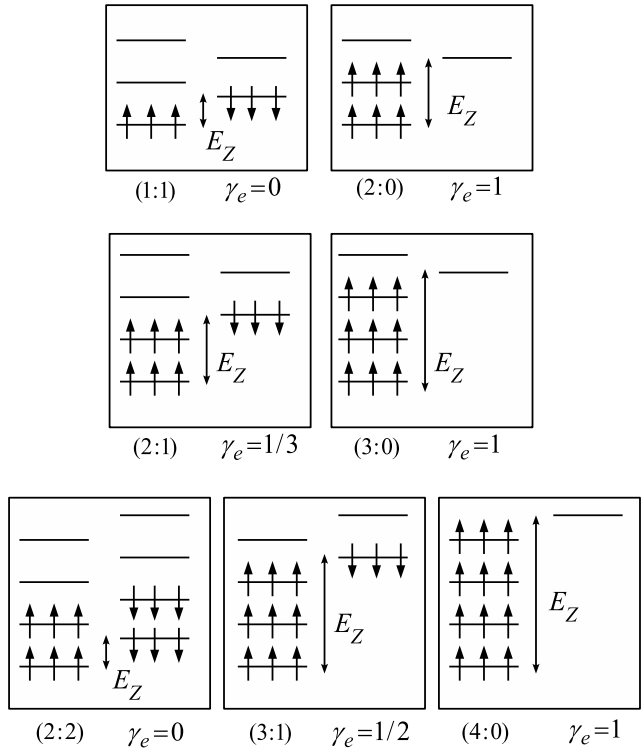}
\caption{A Summary of the qualitative predictions of CF theory (modified from a similar figure in Ref.~\onlinecite{park2001}). The figure shows the filling of effective LLs with spin-up and spin-down composite fermions. Different cases are labelled by their quantum numbers $n_{\uparrow}$ and $n_{\downarrow}$,  the number of filled spin-up and spin-down LLs, e.g., (1:1) denotes $n_{\uparrow}=1$ and $n_{\downarrow}=1$. The degree of spin polarization $\gamma_e$ is calculated using Eq.~(\ref{degreeofspin}). }
\label{qualitativepredictions}
\end{figure}

\end{widetext}

A description of the possible ground-states is summarized in Fig. \ref{qualitativepredictions}. If the Zeeman energy is sufficiently large, then it is expected that the spin will be fully polarized, and so in high magnetic fields the system is pictured by the rightmost diagrams in Fig. \ref{qualitativepredictions}. We then reduce the applied field, keeping the filling factor fixed by lowering the electrons density at the same rate. When the critical field $B^{\mbox{\tiny crit}}$ is reached, then it is energetically favorable for a transition to one of the states with lower net spin polarization, pictured in the diagrams successively to the left in Fig. \ref{qualitativepredictions}. This is energetically favorable as long as the difference in Coulomb energy between the two ground-state configurations compensates for the increase in Zeeman energy due to the spin depolarization.

Comparing these qualitative predictions to the experimental results presented in Ref.~\onlinecite{kukushkin1999}, we find that the predictions for the number of transitions and for the degrees of spin polarizations are broadly correct as a first approximation. In practice, the transitions are somewhat broadened and it is apparent that there are some small second-order plateau occurring between the main transitions. These effects are not well understood and so presently  we shall only focus on the leading-order effects.  Further, we shall assume that the experiment is an observation of ground-state quantum Hall behavior.

\subsection{Trial Wavefunctions for the LLL}
\label{trialwavefunctions}

In order to study the bulk properties of a quantum Hall ground-state we must choose a geometry which eliminates boundary effects. We have chosen to use the spherical geometry for this purpose, i.e., we shall study various finite-sized FQH states existing on the surface of a sphere. To effect a magnetic field perpendicular to the surface of a sphere we must place a magnetic monopole at its center such that the electrons see $N_{\Phi}$ flux quanta. As we increase the system size the total flux increases, but to fix the flux density at the surface of the sphere, the radius of the sphere must also increase. On extrapolation to the thermodynamic limit, therefore, a flat geometry is recovered and the edge effects are eliminated.

We shall now describe how CF states can be expressed as co-ordinate wavefunctions in the sphere geometry, in terms of the spinor co-ordinates $u_i$ and $v_i$ .  (See Appendix \ref{geometry} for a more detailed discussion of the sphere geometry.)

\subsubsection*{Spin polarized CFs}

The CF wavefunction describing interacting electrons in magnetic field $B$ can be succinctly expressed as
\begin{equation}
\label{CFwavefunction}
\psi_{p,n} = \hat{P}_{\mbox{\small LLL}} \left[\Phi^{2p}_0 \Phi_{n}\right],
\end{equation}
where
\[
\Phi_0 = \prod_{i<j} \left(u_i v_j -u_j v_i\right)
\]
and where $\hat{P}_{\mbox{\small LLL}}$ denotes projection onto the lowest Landau level (LLL). $\Phi_{n}$ is a Slater determinant of non-interacting single-fermion wavefunctions with an effective magnetic flux which we denote $2Q$  (i.e., the noninteracting fermions see $2Q$ flux quanta). The quantity $Q$ is known as the effective monopole strength: it can take positive or negative integer or half-integer values. \cite{jainbook} The corresponding effective magnetic field thus can be aligned either parallel or antiparallel to the real magnetic field. The intuition is that $\Phi_{n}$ represents an integer quantum Hall wavefunction for composite fermions in an effective magnetic field and with effective filling $n = \lim_{N \to \infty} N /(2Q)$, i.e., the number of occupied effective LLs is an integer $n$. For a finite sized system, $N /(2Q)$ may be slightly shifted from its thermodynamic value. CF wavefunctions are often simply denoted as $^{2p} \mbox{CF}_n$ or $^{2p} \mbox{CF}_{-n}$ where the sign corresponds to the sign of the effective field and $n$ is now a positive integer. \cite{jainbook} We shall adopt this nomenclature for the remainder of our discussion. The CF states occur at filling factors given in Eq.~(\ref{fillingfactor}). We shall always consider non-interacting CFs.

Practically, the CF states are constructed as follows: for a system of $N$ spin polarized electrons in the spherical geometry filling $n$ CF LLs, the effective monopole strength is given by
\begin{equation} \label{eq:Qvalue}
Q=\pm \frac{N-n^2}{2n},
\end{equation}
with the sign depending on the sign of $B_{\mbox {\tiny eff}}$. The magnetic flux experienced by the electrons due to the magnetic field $B$ is then given by $N_\Phi = 2p(N-1) + 2Q$ [from Eq.~(\ref{CFwavefunction})]. In the sphere geometry, the single-particle CF eigenfunctions are the set of what are called \emph{monopole harmonics}. \cite{jainbook}  These monopole harmonics are eigenfunctions of the effective LL with eigenvalue $n' = 0,....n-1$, of the orbital angular momentum with eigenvalue $l=|Q|, |Q|+1, ... |Q|+n'$, and of the $z$ component of orbital angular momentum with eigenvalue $m=-l,-l+1,...,l$. It is simple to check that given Eq.~(\ref{eq:Qvalue}) the total number of single particle states is $\sum_{i=0}^{n-1} \left({2(|Q|+i)+1}\right) = N$. For $Q<0$, the monopole harmonics are of the following form \cite{jainbook, moller2005}:
\begin{widetext}
\[
Y^{Q<0}_{n',m} (u_i,v_i) = (-1)^{n'} M_{Q,n',m} (u_i^*)^{-Q+m} (v_i^{*})^{-Q-m} \sum_{s=0}^{n'}(-1)^{s}\left( {\begin{array}{*{20}{c}}{n'}  \\s  \\
\end{array}} \right)\left( {\begin{array}{*{20}{c}}
   {2\left| Q \right| + n'}  \\
   {\left| Q \right| + m + s}  \\
\end{array}} \right) (u_i^{*} u_i)^{s} (v_i^{*} v_i)^{n'-s},
\]
where $i$ indicates the particle number and will run from 1 to the total number of particles $N$. Here $M_{Q,n',m}$ is an unimportant normalization factor. We can then write the Slater determinant as
\[
\Phi_{n} = \det \left[ Y^{Q}_{n',m} (u_i,v_i)\right] = \left| \begin{array}{ccc}  Y^{Q}_{0,-|Q|} (u_1,v_1) & \ldots &  Y^{Q}_{0,-|Q|} (u_N,v_N)
 \\ \vdots & & \vdots \\ Y^{Q}_{0,|Q|} (u_1,v_1) & \ldots &  Y^{Q}_{0,|Q|} (u_N,v_N)
\\ Y^{Q}_{1,-|Q|-1} (u_1,v_1) & \ldots &  Y^{Q}_{1,-|Q|-1} (u_N,v_N) \\ \vdots & & \vdots \\ Y^{Q}_{n-1,|Q|+n-1} (u_1,v_1) & \ldots &  Y^{Q}_{n-1,|Q|+n-1} (u_N,v_N) \end{array}\right|.
\]
\end{widetext}

In order to implement the LLL projection required by Eq.~(\ref{CFwavefunction}), we follow the method introduced by Jain and Kamilla \cite{jain1997}, and then extended for the case of a negative effective field by M\"{o}ller and Simon \cite{moller2005}. From a computational perspective, for large system sizes, the LLL projection of the Slater determinant is completely impractical (see Ref.~\onlinecite{jainbook} for details of the LLL projection). The issue can be circumvented by moving the Jastrow factor inside the determinant and then applying $\hat{P}_{\mbox{\small LLL}}$ to each of the resulting matrix elements first, before calculating the determinant:
\begin{equation}
\label{CFprojected}
\psi_{p,n} = \det \left[ \hat{P}_{\mbox{\small LLL}} \left(  Y^{Q}_{n',m}(u_i,v_i)  J_i^p  \right) \right],
\end{equation}
where
\[
J_i = \prod_{j \ne i} \left( u_i v_j - u_j v_i \right).
\]

Although the result of this procedure is not mathematically identical to Eq.~(\ref{CFwavefunction}), the resulting trial wavefunction would nevertheless describe fermions in the LLL, and the two prescriptions have been found to be extremely similar in cases where they can be compared.  An expression for 
\[  {\hat Y}^{Q}_{l,m}(u_i,v_i) J_i  \equiv \,\, 
\hat{P}_{\mbox{\small LLL}}  \left( Y^{Q}_{l,m}(u_i,v_i)  J_i^p \right)
 \] 
in negative effective field was derived in Ref.~\onlinecite{moller2005} and is given by
\begin{widetext}
\[
\hat{Y}^{Q}_{n',m}(u_i,v_i) \propto \sum_{s=0}^{n'} (-1)^{s}\left( {\begin{array}{*{20}{c}}{n'}  \\s  \\
\end{array}} \right)\left( {\begin{array}{*{20}{c}}
   {2\left| Q \right| + n'}  \\
   {\left| Q \right| + m + s}  \\
\end{array}} \right) u^{s}_i v^{n'-s}_i \left( \frac{\partial}{\partial u_i}  \right)^{|Q|+m+s} \left( \frac{\partial}{\partial v_i} \right)^{|Q|-m+n'-s}.
\]
\end{widetext}
Using this result, Eq.~(\ref{CFprojected}) for the CF wavefunction can then be rewritten as
\begin{equation}
\label{CFfinal}
^{2p} \mbox{CF}_{-n} \equiv \psi_{p,n} = \det \left[ \hat{Y}^{Q}_{n',m}(u_i,v_i)  J^p_i  \right],
\end{equation}
In Appendix \ref{CFalgorithm} we discuss an efficient technique for numerically evaluating such wavefunctions.

\subsubsection*{CFs with spin}

If the CFs have a spin degree of freedom, then we must associate a spin degree of freedom with the single-particle monopole harmonic wavefunctions. Let us say we have $N$ CFs with one of two possible spin polarizations,  that is, $N_\uparrow$ spin-up CFs and $N_\downarrow$ spin-down CFs. These CFs can then independently occupy a number $n_{\uparrow}$ and $n_{\downarrow}$ of spin-up or spin-down effective LLs. The value of $p$ is independent of the spin degree of freedom, since it is not involved in the single particle monopole harmonic functions. The spin CF wavefunctions are of the general form:
\begin{equation}
\label{CFspin}
\psi_{p,(n_{\uparrow}, n_{\downarrow})} = \hat{P}_{\mbox{\small LLL}} \left[\Phi^{2p}_0 \Phi_{n_{\uparrow}} \Phi_{n_{\downarrow}}\right],
\end{equation}
where the Slater determinants $ \Phi_{n_{\uparrow}} $ and $\Phi_{n_{\downarrow}}$ are formed from monopole harmonics with the following structures:
\[
Y^{Q}_{n'_{\uparrow},m_{\uparrow}} (u_i,v_i) \, \otimes \, \left| \uparrow \right\rangle \,\,\, , \,\,\, Y^{Q}_{n'_{\downarrow},m_{\downarrow}} (u_i,v_i) \, \otimes \, \left| \downarrow \right\rangle.
\]
The effective monopole strength is now given by:
\begin{equation}
\label{effectiveflux}
Q= \pm \frac{N_{\uparrow}-n_{\uparrow}^2}{2n_{\uparrow}}= \pm \frac{N_{\downarrow}-n_{\downarrow}^2}{2n_{\downarrow}} .
\end{equation}
The possible eigenvalues of the monopole harmonics are now $n'_{\uparrow} = 0,...,n_{\uparrow}$, $l_{\uparrow} =|Q|, ...|Q| + n'_{\uparrow}  $ and $m_{\uparrow} = -l_{\uparrow}, ... l_{\uparrow}$, and similarly for the spin-down versions. Once we construct a Slater determinant of such states we can factor out the antisymmetric spin part of the wavefunction, and we only need to specify the spatial part \cite{jainbook}.

In accordance with Jain's notation, \cite{jainbook} we denote the series of spin un-polarized CF wavefunctions by $^{2p} \mbox{CF}_{(n_{\uparrow},n_{\downarrow})}$ or by $^{2p} \mbox{CF}_{(-n_{\uparrow},-n_{\downarrow})}$ if the effective field is antiparallel to the magnetic field, with $n_{\uparrow}$ and $n_{\downarrow}$ being positive integers (it is not possible to have positive $B_{\mbox {\tiny eff}}$ for one spin species and negative $B_{\mbox {\tiny eff}}$ for the other). The filling factor of the spin-dependent CF states is again given by Eq.~(\ref{fillingfactor})but now with $n=n_{\uparrow}+n_{\downarrow}$.

The final step is to project the wavefunctions onto the LLL; the final form of the spatial part of the spin-dependent CF wavefunction is, thus,
\begin{widetext}
\begin{equation}
\label{CFspinfinal}
^{2p} \mbox{CF}_{(-n_{\uparrow},-n_{\downarrow})} \equiv \psi_{p,(n_{\uparrow}, n_{\downarrow})}= \det \left[ \hat{Y}^{Q}_{n'_{\uparrow},m_{\uparrow}}(u_i,v_i)  J_i  \right] \det \left[ \hat{Y}^{Q}_{n'_{\downarrow},m_{\downarrow}}(u_i,v_i)  J_i  \right],
\end{equation}
\end{widetext}
where the $i$ index runs from 1 to $N_{\uparrow}$ in the first determinant and from $N_{\uparrow}+1$ to $N_{\downarrow}$ in the second. \cite{note1} Note that the $J_i$ function  is exactly as in Eq.~(\ref{CFfinal}) (i.e., it is a product over all spin-up and spin-down particles) and so in Eq.~(\ref{CFspinfinal}) each matrix element in each determinant depends on the co-ordinates of all of the particles.



\section{Results of Monte Carlo Simulations}
\label{results}

In this section we shall present the results of our calculations of the Coulomb energy per electron associated with the seven principle spin CF  trial ground-state wavefunctions in negative effective field: $^2 CF_{-4}$, $^2 CF_{(-3,-1)}$, $^2CF_{(-2,-2)}$, $^2 CF_{-3}$, $^2 CF_{(-2,-1)}$, $^2CF_{-2}$, and $^2CF_{(-1,-1)}$. Using these results we shall deduce some quantitative predictions of CF theory for the critical Zeeman energy per electron at which transitions between different spin states occur.

\subsection{Coulomb Energy}

Our first calculation is the Coulomb energy per electron, $E_C$, of the CF trial ground-state wavefunctions. In general, the Coulomb energy associated with a state described by a wavefunction $\psi$ is calculated using
\begin{equation}
\label{groundstateenergy}
\left\langle \psi | V | \psi \right\rangle = \int d {\mathbf r}_1 ...d {\mathbf r}_N \left| \psi \right|^2 V\left( {\mathbf r}_1, ... {\mathbf r}_N\right),
\end{equation}
where $V$ is the Coulomb potential (in units of $e^2 / \epsilon l_0$):
\[
V = \sum_{i<j}^{N} \frac{1}{R_{ij}} +V_{\mbox{\tiny BG}}.
\]
Here $R_{ij}$ is the distance between pairs of particles in the sphere geometry. Throughout this paper we use the chord distance convention, $R_{ij}^{\mbox{\tiny Chord}} = 2R_S \left| u_i v_j - v_j u_i\right|$, where $R_S$ is the radius of the sphere. $V_{\mbox{\tiny BG}}$ is the potential due to a uniformly changed positive background (we need to put this in so the overall system is electrically neutral), and it is given by the self-energy of a uniformly charged sphere of charge $+Ne$ plus the electrostatic energy between the electrons and the uniformly charged sphere. Using the chord distance measure, $V_{\mbox{\tiny BG}}= -\frac{N^2}{2 R_S}$ (see Ref.~\onlinecite{morf1987}).

One can numerically evaluate integrals of the form given in Eq.~(\ref{groundstateenergy}) using a Metropolis Monte Carlo procedure (see Appendix \ref{montecarlo} for a brief description of how we implemented the Metropolis algorithm). The principal difficulty in using the Metropolis algorithm for our calculation is that the value of the wavefunction must be re-calculated for many millions of different sample particle configurations and, for CF trial wavefunctions, this process can be very computationally demanding. In order to make the procedure viable we require an efficient algorithm with which to evaluate the wavefunctions. A previously described algorithm\cite{moller2005} to evaluate CF wavefunctions for the negative effective field case (see Appendix \ref{CFalgorithm}) is most computationally efficient for larger values of the effective LL number $n$ and very inefficient for the smallest values of $n$, e.g $n=2$ or $n_{\uparrow}$ or $n_{\downarrow}=1$ or 2. For the purposes of studying the spin CF states we have seen that we often need to consider small values of $n_{\uparrow}$ or ${n_\downarrow}$ and, consequently, until now, accurate calculation of energies for the spin CF states has not been computationally feasible. We were able to design an alternative algorithm which is \emph{most} efficient for smallest $n$ and less efficient for larger values of $n$. We shall discuss the details of our algorithm in Appendix \ref{CFalgorithm}. Using this new algorithm enabled us to calculate all of the results presented in this paper.

The results of our calculation of the Coulomb energy per electron of the seven CF trial ground-state wavefunctions of interest are presented in Table \ref{energy};  graphs of the extrapolations to the thermodynamic limit are presented in Fig. \ref{energygraph}.



\subsection{2nd Landau Level Coulomb Interaction}

An interesting consideration is to model equivalent wavefunctions in the 2nd LL, that is, the analogous states occurring at filling factor $\nu = \nu_{\mbox{\tiny LLL}} +2$. One approach would be to construct explicit wavefunctions at this new filling factor, however, the CF wavefunctions in anything other than the LLL are difficult to evaluate. A more efficient alternative is possible, however: the problem of electrons interacting via the Coulomb interaction in a higher Landau level is mathematically equivalent to the problem of electrons in the LLL interacting via an effective potential $V^{\mbox{\tiny eff}} (r)$. An appropriate form for the effective potential is derived in Ref.~\onlinecite{toke2005}, although, strictly speaking, this result was derived only for the disc geometry. However, since we will take the thermodynamic limit at the end of the calculation, we can still use the same potential in the sphere geometry since the disc and the sphere geometries are the same in the thermodynamic limit. The effective potential is given by
\begin{equation}
\label{effectivepotential}
V^{\mbox{\tiny eff}}(r) = \frac{1}{r} +\sum\limits_{i=0}^{6} c_i r^i e^{-r},
\end{equation}
with $r$ in units of the magnetic length.

The values of the coefficients $c_i$ appearing here are deduced by equating the Haldane pseudopotential coefficients \cite{haldane1983} of the effective potential in the LLL, namely
\[
V^{\mbox{\tiny eff}}_m =  \frac{1}{2^{2m+1}m!} \int_0^{\infty} r dr V^{\mbox{\tiny eff}}(r) r^{2m} e^{-r^2 / 4},
\]
to the pseudopotential coefficients of the Coulomb interaction in the second LL, namely
\[
V^1_m = \int_0^{\infty} q dq  \tilde V (q) \left(L_1\left( \frac{q^2}{2} \right)\right)^2 L_m \left(q^2\right) e^{-q^2},
\]
where
\[
\tilde V (q) = \int_0^{\infty} r dr V(r) J_0 (qr).
\]
In these expressions, $L_m$ are Laguerre  polynomials and $J_0$ are Bessel functions. It is sufficient to work with only the first seven coefficients and we have checked that the addition of more coefficients does not change the result significantly. In order to determine the coefficients $c_0$ to $c_6$ in Eq.~(\ref{effectivepotential}) we must equate $V^{\mbox{\tiny eff}}_m$ to $V^1_m$ for $m=1,3,5,\ldots,13$ (since the pseudopotentials vanish for $m=0,2,4,\ldots $ etc.). Appropriate values of $c_i$ are given in Table \ref{valuesofcoefficients}, \cite{note2} and we note that these differ very slightly from the equivalent values presented in Ref.~\onlinecite{toke2005}. The effective Coulomb energies due to the modified interaction potential for the 2nd LL are given alongside the LLL values in Table \ref{energy}.

\begin{table}[H]
\centering
\begin{tabular}{c||c}
\hline \hline
\multicolumn{1}{c||}{Coefficient} & Value\\
\hline \hline
\multicolumn{1}{c||}{$c_0$} & $-50.36597363$ \\
\multicolumn{1}{c||}{$c_1$} & $87.38179510$ \\
\multicolumn{1}{c||}{$c_2$} & $-56.08455086$ \\
\multicolumn{1}{c||}{$c_3$} & $17.76579124$ \\
\multicolumn{1}{c||}{$c_4$} & $-2.971636200$ \\
\multicolumn{1}{c||}{$c_5$} & $0.2513169758$ \\
\multicolumn{1}{c||}{$c_6$} & $-0.008434843187$ \\
\hline \hline
\end{tabular}
\caption{Values for the coefficients in Eq.~(\ref{effectivepotential}). We note that these differ very slightly from the equivalent values presented in Ref.~\onlinecite{toke2005}.}
\label{valuesofcoefficients}
\end{table}

\begin{widetext}

\begin{figure}[H]
\includegraphics[trim = 10mm 10mm 10mm 10mm, clip, width=\textwidth]{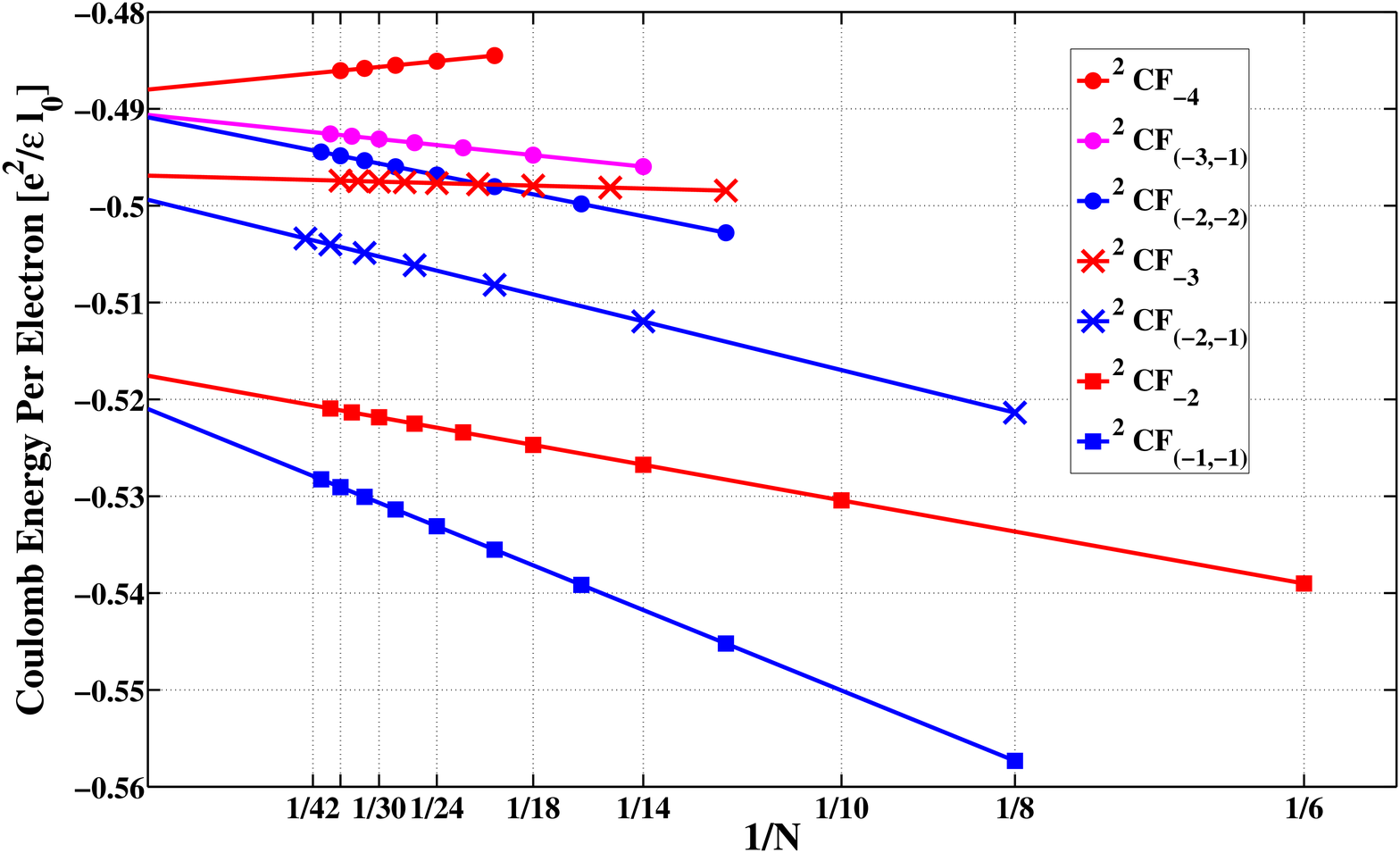}
\caption{Extrapolations to the thermodynamic limit for the Coulomb energies per electron of various CF trial ground-state wavefunctions in the LLL. Only the linear extrapolations in 1/N are shown. Error bars are smaller than the data markers and so are not plotted.}
\label{energygraph}
\end{figure}

\pagebreak
\newpage

\begin{table}[H]
\begin{center}
\subfloat[$^2 CF_{-4}$ State]{\begin{tabular}[t]{c||c||c||c}
\hline \hline
System Size & $N_\Phi$ & Energy (LLL) & Energy (2nd LL) \\
\hline \hline
20 & 37 & $-0.48453(2)$ & $0.5773(1)$\\
24 & 44 & $-0.48508(2)$ & $0.5844(1)$\\
28 & 51 & $-0.48548(2)$ & $0.58980(9)$\\
32 & 58 & $-0.48575(2)$ & $0.59386(9)$\\
36 & 65 & $-0.48614(4)$ & $0.5976(2)$\\
$\infty^1$ & & $-0.4880(1)$ & $0.6222(7)$\\
$\infty^2$ & & $-0.4887(8)$ & $0.629(2)$\\
$\infty^3$ & & $-0.4881(2)$ & $0.623(1)$\\
$\infty^4$ & & $-0.4883(5)$ & $0.625(2)$\\
\hline \hline
\end{tabular}}
\\
\subfloat[$^2 CF_{(-3,-1)}$ State]{
\begin{tabular}[t]{c||c||c||c}
\hline \hline
System Size & $N_\Phi$ & Energy (LLL) & Energy (2nd LL) \\
\hline \hline
14 & 25 & $-0.49595(4)$ & $0.6166(2)$\\
18 & 32 & $-0.49483(3)$ & $0.6201(2)$\\
22 & 39 & $-0.49398(3)$ & $0.6224(1)$\\
26 & 46 & $-0.49353(3)$ & $0.6240(1)$\\
30 & 53 & $-0.49309(4)$ & $0.6246(2)$\\
34 & 60 & $-0.49277(4)$ & $0.6253(2)$\\
38 & 67 & $-0.49266(5)$ & $0.6263(2)$\\
$\infty^1$ & & $-0.49062(6)$ & $0.6318(3)$\\
$\infty^2$ & & $-0.4906(2)$ & $0.6305(9)$\\
$\infty^3$ & & $-0.49059(9)$& $0.6315(4)$\\
$\infty^4$ & & $-0.4907(1)$ & $0.6311(5)$\\
\hline \hline
\end{tabular}}
\qquad \qquad \qquad
\subfloat[$^2CF_{(-2,-2)}$ State]{\begin{tabular}[t]{c||c||c||c}
\hline \hline
System Size & $N_\Phi$ & Energy (LLL) & Energy (2nd LL) \\
\hline \hline
12 & 21 & $-0.50276(3)$ & $0.6430(1)$\\
16 & 28 &  $-0.49986(3)$ & $0.6401(1)$\\
20 & 35 &  $-0.49809(3)$ & $0.6388(1)$\\
24 & 42 &  $-0.49677(3)$ & $0.6375(1)$\\
28 & 49 &  $-0.49597(5)$ & $0.6366(2)$\\
32 & 56 &  $-0.49538(5)$ & $0.6365(2)$\\
36 & 63 &  $-0.49488(4)$ & $0.6355(2)$\\
40 & 70 &  $-0.49442(4)$ & $0.6354(2)$\\
$\infty^1$ &  &  $-0.49088(5)$ & $0.6321(2)$\\
$\infty^2$ &  &  $-0.4908(2)$ & $0.6321(5)$\\
$\infty^3$ &  &  $-0.49082(6)$ & $0.6321(2)$\\
$\infty^4$ & & $-0.4908(1)$ & $0.6319(3)$\\
\hline \hline
\end{tabular}}
\\

\subfloat[$^2 CF_{-3}$ State]{\begin{tabular}[t]{c||c||c||c}
\hline \hline
System Size & $N_\Phi$ & Energy (LLL) & Energy (2nd LL) \\
\hline \hline
12 & 21 & $-0.49843(3)$ & $0.6247(1)$\\
15 & 26 & $-0.49810(3)$ & $0.6325(1)$\\
18 & 31 & $-0.49801(3)$ & $0.6382(1)$\\
21 & 36 & $-0.49781(2)$ & $0.6420(1)$\\
24 & 41 & $-0.49765(2)$ & $0.6448(1)$\\
27 & 46 & $-0.49756(2)$ & $0.6472(1)$\\
30 & 51 & $-0.49754(4)$ & $0.6494(2)$\\
33 & 56 & $-0.49745(4)$ & $0.6509(2)$\\
36 & 61 & $-0.49741(4)$ & $0.6522(2)$\\
$\infty^1$ & &  $-0.49690(4)$ & $0.6657(2)$\\
$\infty^2$ & &  $-0.4968(1)$ & $0.6672(5)$\\
$\infty^3$ & &  $-0.49688(6)$ & $0.6660(2)$\\
$\infty^4$ & & $-0.49679(5)$ & $0.6661(3)$\\
\hline \hline
\end{tabular}}
\qquad \qquad \qquad
\subfloat[$^2 CF_{(-2,-1)}$ State]{
\begin{tabular}[t]{c||c||c||c}
\hline \hline
System Size & $N_\Phi$ & Energy (LLL) & Energy (2nd LL) \\
\hline \hline
8 & 13 & $-0.52139(5)$ & $0.7091(2)$\\
14 & 23 & $-0.51188(4)$ & $0.6930(2)$\\
20 & 33 & $-0.50821(4)$ & $0.6868(2)$\\
26 & 43 & $-0.50611(4)$ & $0.6833(2)$\\
32 & 53 & $-0.50488(4)$ & $0.6814(1)$\\
38 & 63 & $-0.50416(6)$ & $0.6805(2)$\\
44 & 73 & $-0.50328(6)$ & $0.6794(2)$\\
$\infty^1$  & & $-0.49938(6)$ & $0.6724(2)$\\
$\infty^2$  & & $-0.4994(2)$ & $0.6733(4)$\\
$\infty^3$  & & $-0.4994(1)$ & $0.6728(3)$\\
$\infty^4$ & & $-0.4994(2)$ & $0.6732(5)$\\
\hline \hline
\end{tabular}}
\\
\end{center}
\caption{Coulomb energy per electron, $E_C$, for trial wavefunctions at filling factors $\nu = 2/3 , 3/5$, and $4/7$, for various system sizes; some of these data are plotted in Fig. \ref{energygraph}. Results are given for the Coulomb interaction in the LLL, and for the modified Coulomb interaction in the 2nd LL. Energies are calculated using chord distance measure in the sphere geometry, as defined in the text, and are stated in units of $e^2 / {\epsilon l_0}$. The extrapolations to the thermodynamic limit were calculated using three different methods, indicated by the superscript 1--4: 1 is with linear extrapolation in $1/N$, weighted by the statistical errors on each data point; 2 is with quadratic  extrapolation in $1/N$, weighted by the statistical errors on each data point; 3 is with a weighted linear extrapolation in $1/N$, excluding the smallest system size; and 4 is with a weighted linear extrapolation in $1/N$ excluding the two smallest system sizes. The different cases are calculated in order to estimate the possible error in the extrapolations. We note that for the $^2 CF_{-4}$, $^2 CF_{-3}$ and $^2 CF_{-2}$ states, our numbers are in good agreement with those obtained in Ref. \onlinecite{moller2005}.\label{energy}}

\end{table}
\begin{table}[H]
\ContinuedFloat
\begin{center}
\subfloat[$^2CF_{-2}$ State]{\begin{tabular}[t]{c||c||c||c}
\hline \hline
System Size & $N_\Phi$ & Energy (LLL) & Energy (2nd LL) \\
\hline \hline
6 & 9 & $-0.53899(4)$ & $0.7716(2)$\\
10 & 15 & $-0.53047(3)$ & $0.7677(2)$\\
14 & 21 & $-0.52675(3)$ & $0.7669(2)$\\
18 & 27 & $-0.52468(3)$ & $0.7667(1)$\\
22 & 33 & $-0.52342(2)$ & $0.7663(1)$\\
26 & 39 & $-0.52257(5)$ & $0.7665(2)$\\
30 & 45 & $-0.52181(5)$ & $0.7663(2)$\\
34 & 51 & $-0.52129(4)$ & $0.7663(2)$\\
38 & 57 & $-0.52096(4)$ & $0.7665(2)$\\
$\infty^1$ &  & $-0.51755(2)$ & $0.7649(3)$\\
$\infty^2$ &  & $-0.51751(5)$ & $0.7666(1)$\\
$\infty^3$ &  & $-0.51753(3)$ & $0.7657(1)$\\
$\infty^4$ & & $-0.51754(5)$ & $0.7660(1)$\\
\hline \hline
\end{tabular}}
\qquad \qquad \qquad
\subfloat[$^2CF_{(-1,-1)}$ State]{\begin{tabular}[t]{c||c||c||c}
\hline \hline
System Size & $N_\Phi$ & Energy (LLL) & Energy (2nd LL) \\
\hline \hline
8  & 11 & $-0.55748(4)$ & $0.8754(2)$\\
12  & 17 & $-0.54506(4)$ & $0.8355(1)$\\
16  & 23 & $-0.53906(3)$ & $0.8172(1)$\\
20 & 29 & $-0.53536(7)$ & $0.8061(1)$\\
24  & 35 & $-0.53309(3)$ & $0.7991(1)$\\
28 & 41 & $-0.53133(5)$ & $0.7942(1)$\\
32 & 47 & $-0.53016(3)$ & $0.7904(1)$\\
36 & 53 & $-0.52914(3)$ & $0.7874(1)$\\
40 & 59 & $-0.52837(3)$ & $0.7850(1)$\\
$\infty^1$ &  & $-0.52102(8)$ & $0.7621(5)$\\
$\infty^2$ &  & $-0.52144(6)$ & $0.7652(3)$\\
$\infty^3$ &  & $-0.52118(5)$ & $0.7633(2)$\\
$\infty^4$ & & $-0.52123(6)$ & $0.7636(2)$\\
\hline \hline
\end{tabular}}
\\
\end{center}
\caption{(Continued)}
\end{table}

\end{widetext}



\subsection{Comparison with Experiment}

At this point we are now ready to calculate the predicted values for the critical Zeeman energy per electron $E^{\mbox{\tiny crit}}_Z$ for transitions and to make comparisons with the experiments. To study each transition, e.g., at filling $\nu=\frac{4}{7}$ , $\gamma_e = 1\, \rightarrow \, \frac{1}{2}$ (for which the CF theory prediction comes from the transition $^2 CF_{-4} \, \rightarrow \, ^2 CF_{(-3,-1)}$), we need to compare the two key energy scales: first, there is the difference in Coulomb energy per electron, $\Delta E_{C}$,  between the two participant ground-state trial wavefunctions; second, there is the difference in Zeeman energy per electron, $\Delta E_Z$, which is due to the difference in the net spin polarization of the two participant ground-state trial wavefunctions. The condition for a transition is that  $\Delta E_{C} = \Delta E_Z $ at the point of transition.

The difference in Coulomb energy per electron, $\Delta E_{C}$, is calculated directly from our results for the Coulomb energy per electron associated with each trial ground-state wavefunction---these differences are listed in Table \ref{criticalzeemanvalues}. The difference in Zeeman energy per electron as we go through a transition is equal to the energy  $E^{\mbox{\tiny crit}}_Z$ to flip a single spin in a magnetic field $B^{\mbox{\tiny crit}}$ multiplied by the proportion of spins $\tau$ which need to be flipped as we go through that transition. In the thermodynamic limit the proportion of flipped spins is $\tau = 1/n$, where $n$ is the total number of filled effective LLs of the CF wavefunctions partaking in the transition (see Appendix \ref{extrastuff}). Thus, at the transition we have $\Delta E_{C} = \Delta E_Z =E^{\mbox{\tiny crit}}_Z/n$ and this leads to the relation
\begin{equation}
E^{\mbox{\tiny crit}}_Z = n \Delta E_{C}.
\label{criticalzeeman}
\end{equation}
Using Eq.~(\ref{criticalzeeman}) we have calculated predicted values for the critical Zeeman energy, and these predictions are listed in Table \ref{criticalzeemanvalues}.

Values of $E^{\mbox{\tiny crit}}_Z$ can also be deduced from the experimental results presented in Refs. \onlinecite{kukushkin1999,du1995}. In Ref.~\onlinecite{kukushkin1999} the values are derived from the measurements of the degree of spin polarization $\gamma_e$ as a function of magnetic field strength at fixed filling factor. Accompanying each transition is a broadened step in the degree of spin polarization. We take the critical field $B^{\mbox{\tiny crit}}$ to occur at the center of the step, and we take the experimental error to be roughly the half-width of the broadening (which corresponds approximately to an error of $\pm$ 0.001 in the Zeeman energy).  In Ref.~\onlinecite{du1995} we infer the critical fields for spin transitions at filling factors $\nu = 2/3, 3/5$ and $4/7$ from the critical fields for spin transitions observed in particle-hole conjugate states occurring at filling factors $\nu = 4/3, 7/5$ and $10/7$. Particle-hole conjugation does not, in principle, affect the Coulomb or Zeeman energy associated with the trial CF wavefunctions and so the predictions for the critical Zeeman energy are identical. The aim of the experiments described in Ref.~\onlinecite{du1995} is to vary the ratio of Zeeman to Coulomb energy keeping the filling factor fixed. In order to achieve this the electron density is kept constant and the applied field $B^{\mbox{\tiny tot}} $ is tilted by an angle $\theta$ from the vertical\cite{tilt} and increased in magnitude simultaneously, thus fixing the component of the field perpendicular to the plane, $B_{\perp} = B^{\mbox{\tiny tot}} \cos \theta$.  The signature for transitions between states of different spin polarization is a peak in the ratio of the longitudinal resistivity at a given filling factor to the longitudinal resistively at filling $\nu=3/2$. We take the critical field $B^{\mbox{\tiny crit}}$ to occur at the center of the peaks and we take the experimental error to be the half-width of the peaks (which corresponds approximately to an error of $\pm$0.001 in the Zeeman energy). Using these values for $B^{\mbox{\tiny crit}}$ we calculate the experimentally measured value of the critical Zeeman energy via $E^{\mbox{\tiny crit}}_Z = g \mu_B B^{\mbox{\tiny crit}}$, and then convert to units of $e^2/\epsilon l_0$ for comparison with our theoretically derived values in Table \ref{criticalzeemanvalues} (for this calculation we take the $g$ factor of GaAs to be $g=-0.44$ and the relative permittivity to be $\epsilon=12.6$).

In Fig. \ref{theoryvsexperiment} we have plotted the predicted and measured values of $E^{\mbox{\tiny crit}}_Z$ given in Table \ref{criticalzeemanvalues} as a function of the parameter $n$. 
\begin{widetext}

\renewcommand\arraystretch{1.2}

\begin{table}[H]
\begin{center}
\begin{tabular}{c||c||c||c||c||c||c}
\hline \hline
Filling & Transition & CF Theory Prediction&  $\Delta E_{C}$ & $E^{\mbox{\tiny crit}}_Z$ (Predicted) & $E^{\mbox{\tiny crit}}_Z$ (Kukushkin \emph{et al}.) & $E^{\mbox{\tiny crit}}_Z$ (Du \emph{et al}.)\\
\hline \hline
$\frac{4}{7}$ & $\gamma_e=1 \, \rightarrow \,  \frac{1}{2}$&  $^2 CF_{-4} \, \rightarrow \, ^2 CF_{(-3,-1)}$ & 0.0027(2) &  0.0107(8) & 0.013(1) & 0.018(1)\\
 $\frac{4}{7}$ & $\gamma_e=\frac{1}{2} \, \rightarrow \, 0$ & $^2 CF_{(-3,-1)} \, \rightarrow \,^2 CF_{(-2,-2)}$& 0.0003(1) & 0.0010(4) & 0.007(1) &  0.010(1) \\
$\frac{3}{5}$ & $\gamma_e=1 \, \rightarrow \,  \frac{1}{3}$ &  $^2 CF_{-3} \, \rightarrow \,^2 CF_{(-2,-1)}$ & 0.0025(1)&  0.0074(3) & 0.012(1)  &  0.017(1)\\
$\frac{2}{3}$ & $\gamma_e=1 \, \rightarrow \,  0$ &   $^2 CF_{-2} \, \rightarrow \,^2 CF_{(-1,-1)}$ & 0.00363(7) &  0.0073(1) & 0.0088(1) &  0.015(1)\\
\hline \hline
\end{tabular}
\end{center}
\caption{For each filling factor in the negative effective field case, the table shows the possible spin-transitions labelled by their degree of spin polarization $\gamma_e$ [defined in Eq. (\ref{degreeofspin})], the difference in Coulomb energy per electron between the two possible ground-state configurations, and the corresponding prediction for the critical Zeeman energy per electron calculated using Eq.~(\ref{criticalzeeman}). We use the extrapolation scheme that gives minimum uncertainty in the extrapolated value. The relevant experimentally derived values for the critical Zeeman energy taken from Kukushkin \emph{et al}. (Ref.~\onlinecite{kukushkin1999}) and from Du \emph{et al}. (Ref.~\onlinecite{du1995}). We explain in the text how the experimental values are deduced. All values are given in units of $e^2/\epsilon l_0$.}
\label{criticalzeemanvalues}
\end{table}

\renewcommand\arraystretch{1.0}

\begin{figure}[H]
\includegraphics[trim = 10mm 10mm 10mm 10mm, clip, width=\textwidth]{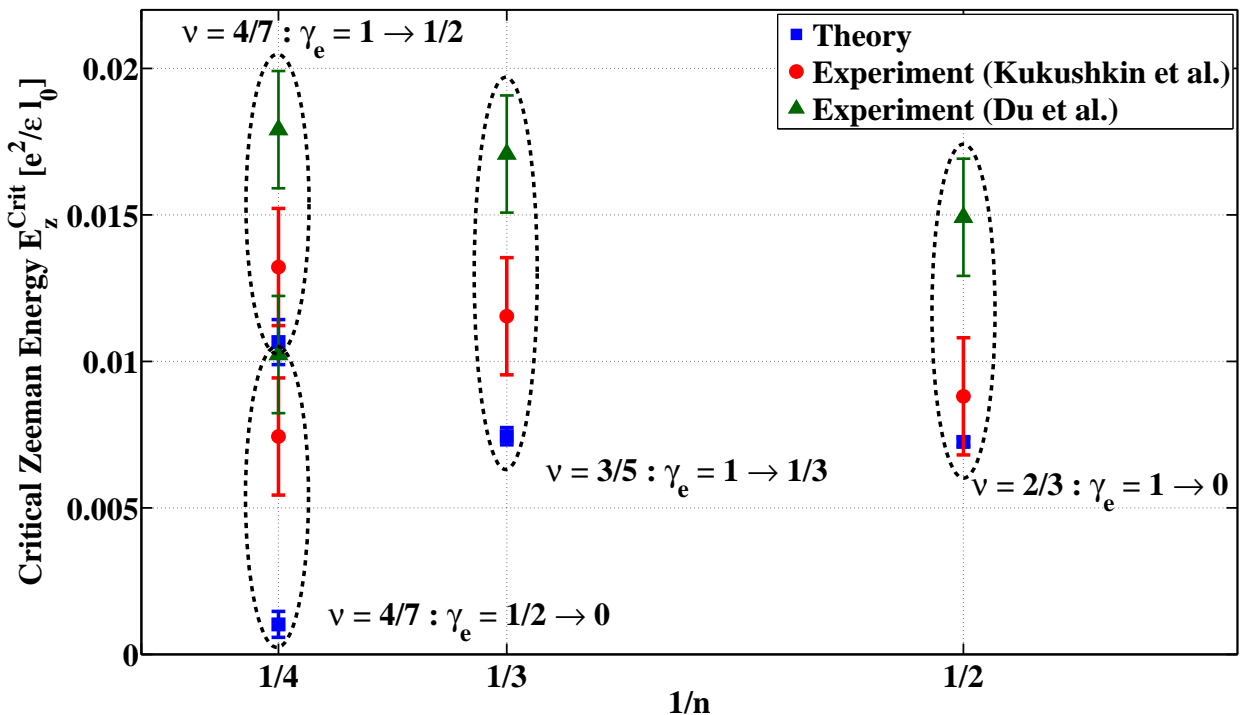}
\caption{Predicted and measured values for the critical Zeeman energy per electron from Table \ref{criticalzeemanvalues} plotted against the reciprocal number of filled effective LLs, $1/n$. Transitions are labelled by their degree of spin polarization $\gamma_e$ [defined in Eq.~(\ref{degreeofspin})]. In the text we explain how the theoretical predictions of the Critical Zeeman per electron are calculated. Experimental values are taken from Kukushkin \emph{et al}. (Ref.~\onlinecite{kukushkin1999}) and from Du \emph{et al}. (Ref.~\onlinecite{du1995}). We explain in the text how the experimental values are deduced. Circled sets of points correspond to the same transition. For the series of $^2 CF_{-n} \, \rightarrow \, ^2 CF_{(-n-1,-1)}$ transitions (i.e., those transitions for which higher value of $\gamma_e$ is 1) there is a trend for the critical Zeeman energy to increase as $n$ increases.  In all cases theory tends to underestimate the critical Zeeman energy.}
\label{theoryvsexperiment}
\end{figure}
\pagebreak
\end{widetext}

\subsection{Finite Thickness Correction}

\begin{widetext}

\begin{figure}[H]
\begin{center}
\subfloat[Interaction energy per electron associated with a selection of the CF ground-states in the thermodynamic limit due to a modified Coulomb potential $V_{\mbox{\tiny FH}} (r,d)$ given in Eq.~(\ref{finitethickness}), plotted as a continuous  function of the thickness parameter $d$. The associated error bars for these extrapolations are not drawn as they are too small to see on the plot.]{
\includegraphics[trim = 10mm 5mm 10mm 10mm, clip, width=0.8\textwidth]{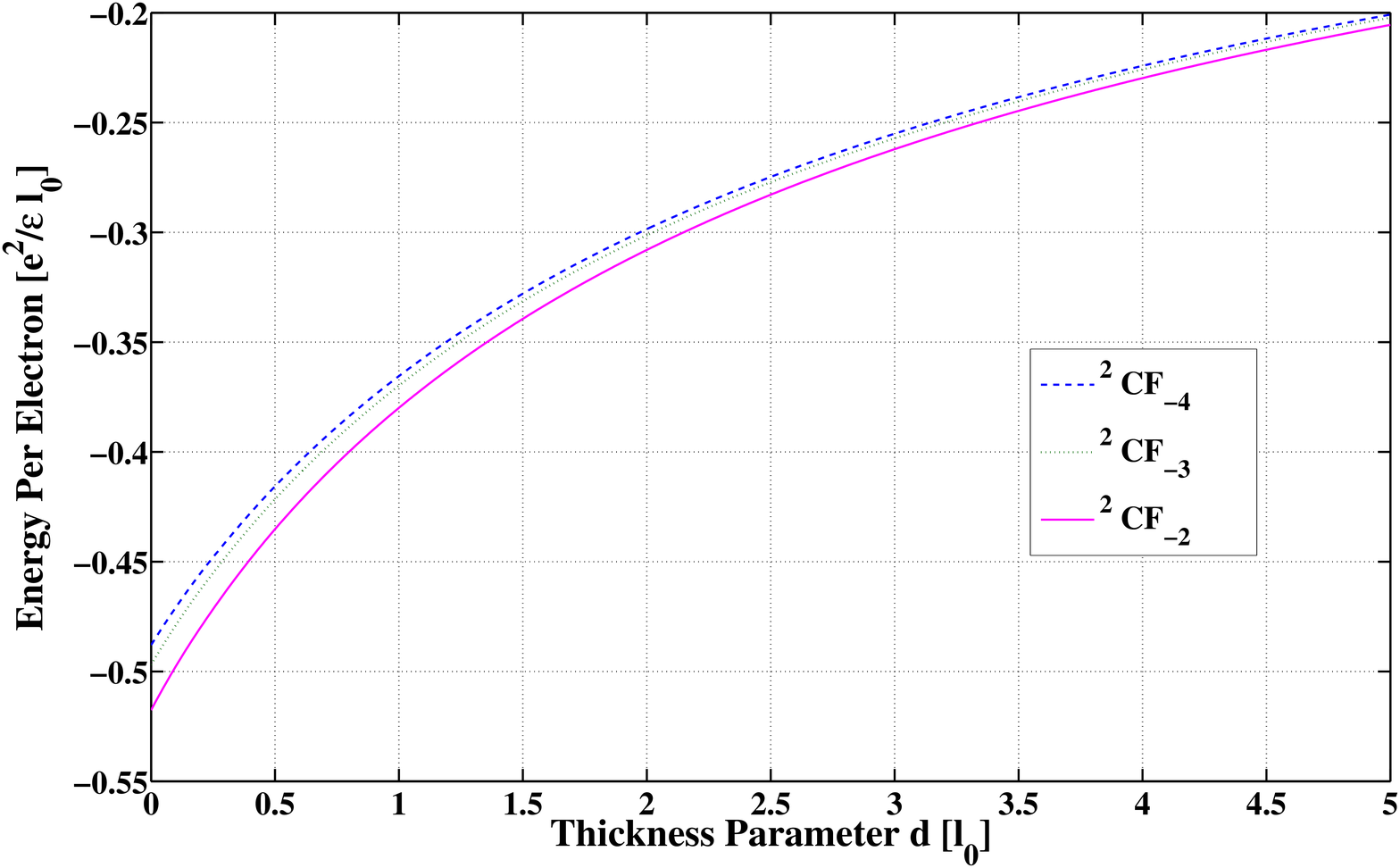}
}
\\
\subfloat[Theoretical predictions for the critical Zeeman energy per electron due to a finite-thickness potential  $V_{\mbox{\tiny FH}} (r,d)$ given in Eq.~(\ref{finitethickness}), plotted as a continuous function of the thickness parameter $d$ for selected CF states. The critical Zeeman energy is calculated using Eq.~(\ref{criticalzeeman}).  Representative error bars  for a selected subset of the data are drawn at intervals on the plot. Note that the data have been smoothed between $d=0.2$ and 0.3 in order to remove a numerical artifact.]
{
\includegraphics[trim = 10mm 5mm 10mm 10mm, clip, width=0.8\textwidth]{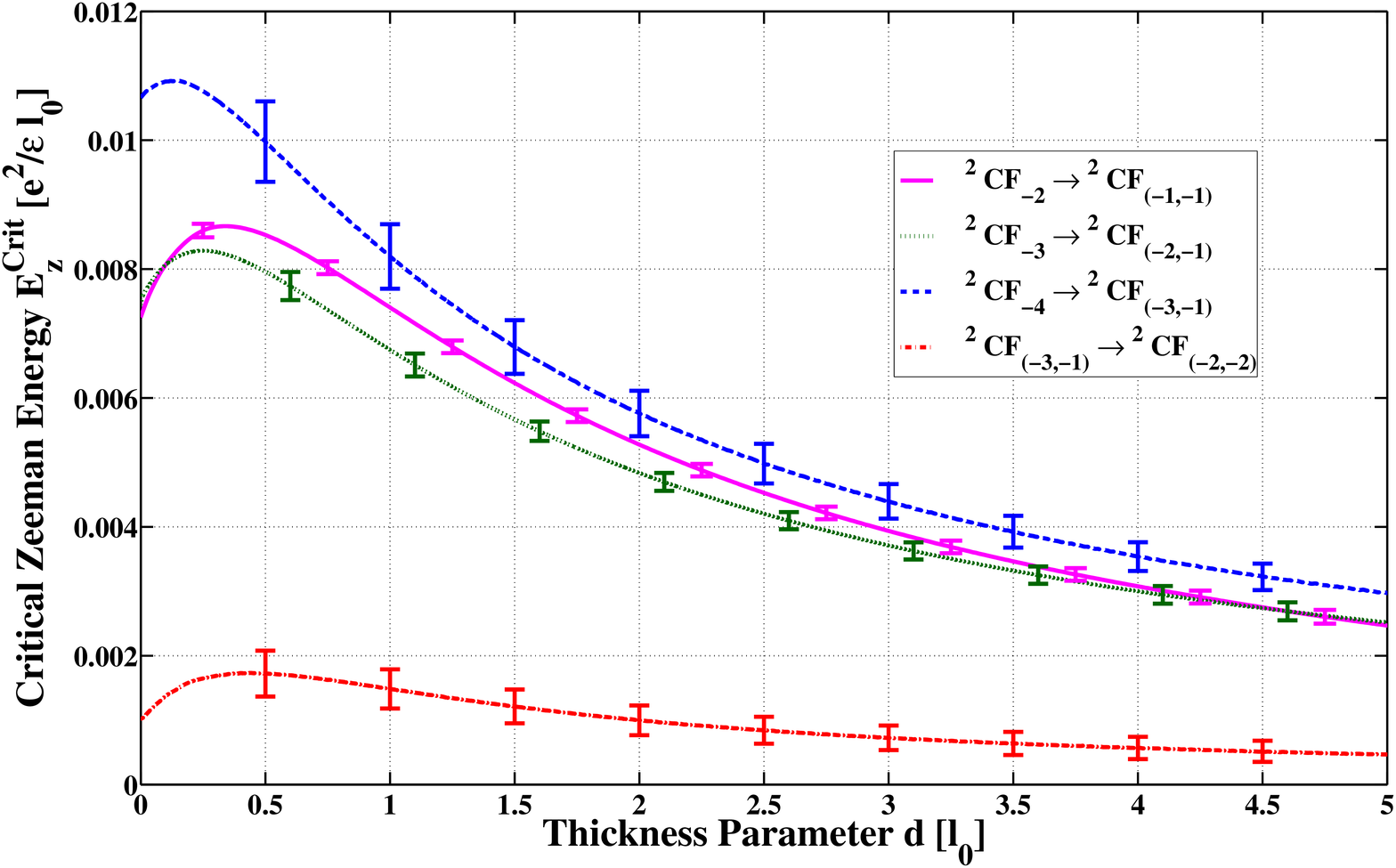}
}
\caption{Graphs showing the impact of the finite thickness correction on our results. The energy is given in units of $e^2/\epsilon l_0$ and the thickness parameter is given in units of $l_0$. To obtain these curves, weighted extrapolations to the thermodynamic limit are calculated for the sets of  data points at finite $N$ for 501 discrete values of the parameter $d$ between 0 and 5 and then the curves are interpolated. We use the extrapolation scheme that gives minimum uncertainty in the extrapolated value  (i.e., possibly with the smallest system size removed from the extrapolation).}
\label{fintewidthgraphs}
\end{center}
\end{figure}

\end{widetext}

In the experimental investigations of the FQHE discussed in Refs. \onlinecite{kukushkin1999,du1995}, the 2D geometry is realized in a GaAs--$\mbox{Al}_x \mbox{Ga}_{1-x}\mbox{As}$ heterojunction setup. Due to the finite width of the potential well in heterojunctions, the geometry cannot be considered perfectly 2D and realistic wavefunctions must have some finite extent in the direction perpendicular to the 2D plane. The impact of the finite thickness correction can be calculated simply by modifying the interaction potential with which we evaluate the ground-state energy. Appropriate modified potentials are discussed, for example in Ref.~\onlinecite{peterson2008}. Here we implement the following potential, known as the Fang-Howard potential, which is applicable to GaAs--$\mbox{Al}_x \mbox{Ga}_{1-x}\mbox{As}$ heterojunctions. Note that this result was derived only for the disc geometry, however in the thermodynamic limit, the result will be correct for the sphere geometry also.

\begin{equation}
\label{finitethickness}
V_{\mbox{\tiny FH}} (r,d) =V'_{\mbox{\tiny BG}} + \int_0^{\infty} dk V_{\mbox{\tiny FH}} (k,d)  J_0 (kr) k,
\end{equation}
with
\[
V_{\mbox{\tiny FH}} (k,d) = \frac{e^2 l_0}{\epsilon} \frac{9}{8k} \frac{24+9kd+(kd)^2}{\left( 3 + kd \right)^3}.
\]
The potential is a function of a thickness parameter $d/l_0$, which characterizes the extent of the wavefunction in the dimension perpendicular to the plane; $d=0$ corresponds to a perfect 2D geometry. Note that the value of the background contribution to the potential, $V'_{\mbox{\tiny BG}}$, will depend on this new potential and will not be the same as for the Coulomb potential case. Using this modified interaction potential we have calculated the ground-state energy in the thermodynamic limit for a number of different values of $d$. Graphs showing the thermodynamic extrapolation of the ground-state energy as a function of the thickness parameter are presented in Fig. \ref{fintewidthgraphs}.

The value of $d$ for a particular GaAs--$\mbox{Al}_x \mbox{Ga}_{1-x}\mbox{As}$ heterojunction can be estimated using known data about the system (see Ref.~\onlinecite{stern1967}):
\begin{equation}
d=\frac{1}{3} \left( \frac{48 \pi m_z e^2 \rho^*}{\kappa_{\mbox{sc}} \hbar^2}\right)^{-\frac{1}{3}},
\label{estimatethickness}
\end{equation}
where $m_z$ is the effective mass of the electron in the GaAs in the direction perpendicular to the plane of the heterojunction, which is $0.063m_e$; $\kappa_{\mbox{sc}}$ is the static dielectric constant of GaAs, which is $12.6$ and $\rho^*$ for an un-doped system is given by $\frac{11}{32}\rho_s$, where $\rho_s$ is the areal density of electrons in the inversion layer. The value must then be expressed in units of the magnetic length $l_0$.

In the experiments of Kukushkin \emph{et al}. (Ref.~\onlinecite{kukushkin1999}) the system is setup such that the ratio of the $\rho_s$ to $B^{\mbox{\tiny crit}}$ is fixed by the filling factor, i.e., $\rho_s = \nu B^{\mbox{\tiny crit}} / \phi_0$  (here $\phi_0$ is the flux quantum $\phi_0 = hc/e$), and, thus, in Eq.~(\ref{estimatethickness}) the value of $d$ is a function of $B^{\mbox{\tiny crit}}$. An additional dependence of $d$ on $B^{\mbox{\tiny crit}}$ enters through the magnetic length. Theoretical predictions for $B^{\mbox{\tiny crit}}$ can be derived from the values of the critical Zeeman energy given in Table \ref{criticalzeemanvalues} or, if we consider a system with non-zero $d$, from the results presented in Fig. \ref{fintewidthgraphs}. Clearly, then, the theoretical prediction for $B^{\mbox{\tiny crit}}$ itself depends on the chosen value of $d$, estimated from Eq.~(\ref{estimatethickness}). For each type of transition we can determine theoretical predictions for the values of $d$ by requiring that the above conditions are self-consistently satisfied. We have tabulated these self-consistent estimates in Table \ref{valuesofd}.

In the experiments of Du \emph{et al}. (Ref.~\onlinecite{du1995}) there are two slight modifications to our estimation of $d$. First, in these tilted field experiments the value of $\rho_s$ is fixed at $1.13~\times10^{11}~\mbox{cm}^{-2}$. Second, the effective magnetic length due to the in-plane field $B^{\mbox{\tiny tot}}$ places an additional limit on the extent of the wavefunction perpendicular to the plane. The in-plane field can be determined by the requirement that the filling factor is fixed and the number density of electrons is known, i.e., we require that ${B_ \bot } = \rho_s \phi_0 / \nu$; the value of the total field  is derived from the prediction of the critical Zeeman energy; the in-plane field  ${B_\parallel }$ is given by ${B_\parallel }=\sqrt{ (B^{\mbox{\tiny tot}})^2 - ({B_ \bot })^2}$. In fact, since our predictions for the critical Zeeman energy are relatively small compared to the values observed in Du's experiment, we find that our predicted values for $B^{\mbox{\tiny crit}}$ are smaller than the values ${B_ \bot }$ consistent with the experimental electron density and filling factor. It is, therefore, not possible in this case to construct a self-consistent value for ${B_\parallel }$. We can, however, use the experimentally observed values of the critical field to make an estimate of the value of $d$ in units of $l_0$ for each transition using Eq.~(\ref{estimatethickness}).

In Table \ref{valuesofd} we present our estimates of $d$ appropriate for each experimental setup described in the preceding two paragraphs and we present the corresponding theoretical predictions for the critical Zeeman energy per electron calculated using the potential described in Eq.~(\ref{finitethickness}). In Fig. \ref{modifiedtheoryvsexperiment} we plot the modified results for the critical Zeeman energy per electron given in  Table \ref{valuesofd} as a function of $n$ as in Fig. \ref{theoryvsexperiment}.

\pagebreak

\begin{widetext}

\renewcommand\arraystretch{1.2}

\begin{table}[H]
\begin{center}
\begin{tabular}{c||c||c||c||c||c||c}
& & & \multicolumn{2}{c||}{Kukushkin \emph{et al}.} & \multicolumn{2}{c}{Du \emph{et al}.} \\
\hline \hline
Filling & Transition & CF Theory Prediction &  $d $ & $E_z^{\mbox{\tiny crit}}$ &  $d$ & $E_z^{\mbox{\tiny crit}}$ \\
\hline \hline
$\frac{4}{7}$ &  $\gamma_e=1 \, \rightarrow \,  \frac{1}{2}$  &$^2 CF_{-4} \, \rightarrow \, ^2 CF_{(-3,-1)}$ &  1.42 & 0.0070(4)& 0.595 &  0.0096(6) \\
$\frac{4}{7}$ &  $\gamma_e=\frac{1}{2} \, \rightarrow \, 0$ & $^2 CF_{(-3,-1)} \, \rightarrow \,^2 CF_{(-2,-2)}$ &  0.865 &  0.0016(3) &  1.20 &  0.0014(3)\\
$\frac{3}{5}$  &  $\gamma_e=1 \, \rightarrow \,  \frac{1}{3}$ & $^2 CF_{-3} \, \rightarrow \,^2 CF_{(-2,-1)}$ &  1.33 & 0.0060(2) & 0.636 &  0.0076(2) \\
 $\frac{2}{3}$ & $\gamma_e=1 \, \rightarrow \,  0$ & $^2 CF_{-2} \, \rightarrow \,^2 CF_{(-1,-1)}$ &1.44 &  0.0063(1) &  0.776 & 0.0078(1) \\
\hline \hline
\end{tabular}
\end{center}
\caption{The table presents estimates of the finite thickness parameter $d$ in units of $l_0$ for different transitions and filling factors in negative effective field. Transitions are labelled by their degree of spin polarization $\gamma_e$ [defined in Eq.~(\ref{degreeofspin})]. The estimates are calculated using Eq.~(\ref{estimatethickness}) and experimental data taken from  Kukushkin \emph{et al}. (Ref.~\onlinecite{kukushkin1999}) or from Du \emph{et al}. (Ref.~\onlinecite{du1995}). The table also presents modified theoretical predictions for the critical Zeeman energy per electron in units of $e^2/\epsilon l_0$ calculated by using the estimates for $d$ with the modified potential given in Eq.~\ref{finitethickness}.}
\label{valuesofd}
\end{table}

\renewcommand\arraystretch{1.0}

\begin{figure}[H]
\includegraphics[trim = 10mm 10mm 10mm 10mm, clip, width=\textwidth]{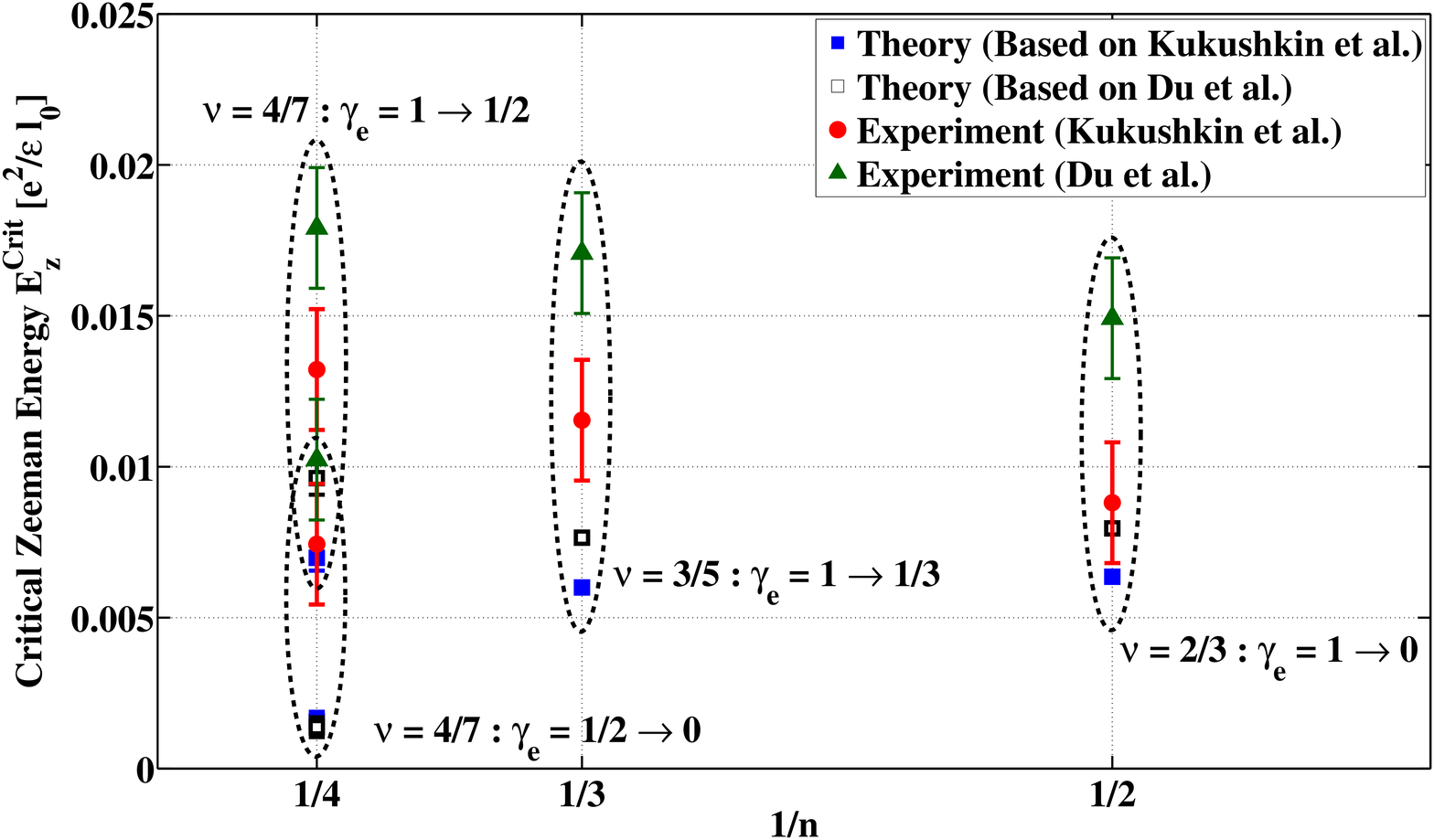}
\caption{Predicted values for the critical Zeeman per electron modified by the inclusion of finite thickness effects (data from Table \ref{valuesofd}) and measured values for the critical Zeeman energy per electron from Table \ref{criticalzeemanvalues} plotted against the reciprocal number of filled effective LLs, $1/n$. Transitions are labelled by their degree of spin polarization $\gamma_e$ [defined in Eq. ~(\ref{degreeofspin})]. In the text we explain how the theoretical predictions of the critical Zeeman energy per electron are calculated, taking into account a finite thickness correction to the potential appropriate to the conditions of the different experimental systems considered. Experimental values are taken from Kukushkin \emph{et al}. (Ref.~\onlinecite{kukushkin1999}) and from Du \emph{et al}. (Ref.~\onlinecite{du1995}). We explain in the text how the experimental values are deduced. Circled sets of points correspond to the same transition.  As in Fig. \ref{theoryvsexperiment}, the theory somewhat underestimates the critical Zeeman energy.}
\label{modifiedtheoryvsexperiment}
\end{figure}

\end{widetext}



\section{Conclusions}
\label{conclusion}

Looking at Table \ref{criticalzeemanvalues},  we conclude that the predictions made by CF theory for the critical Zeeman energy agree moderately well with the experimental values: we find a good agreement for the $^2 CF_{-2}$ to $^2 CF_{(-1,-1)}$ transition, but we find that the predictions for the $^2 CF_{-3}$ to $^2 CF_{(-2,-1)}$, the $^2 CF_{-4}$ to $^2 CF_{(-3,-1)}$, and the $^2 CF_{(-3,-1)}$ to $^2 CF_{(-2,-2)}$ transitions are, respectively, at least a factor of 1.6, 1.2 and 7 away from the experimental values. Looking at Fig. \ref{theoryvsexperiment}, we notice a trend in both sets of experimental data for the critical Zeeman energy to increase with increasing $n$ (excluding the $^2 CF_{(-3,-1)}$ to $^2 CF_{(-2,-2)}$ transition). This trend is also present in the theoretical predictions of CF theory. Compared with the analogous set of results for the positive effective field case (filling factors $\nu=2/5 , 3/7$ and $4/9$, see Ref. \onlinecite{park2001}), those results tend to agree much better with experiment for the series of $^2 CF_{n} \, \rightarrow \, ^2 CF_{(n-1,1)}$; however, there remains a similar large discrepancy with experiment for the $^2 CF_{(3,1)}$ to $^2 CF_{(2,2)}$ transition at filling $4/9$. 

The potential impact of the finite thickness correction, for small values of the thickness parameter ($d<0.5$), is to increase the differences in the predicted critical Zeeman energy per particle between the adjacent states typically by 20--30\%. For larger values of the thickness parameter ($d>0.5$ ) the predicted critical Zeeman energy per particle decreases.  Based on our estimates of the value of $d$, given in Table \ref{valuesofd}, we conclude the the effect of the finite thickness correction would in fact be to lower our predicted values for the critical Zeeman energy per electron. We do however, correctly predict that the critical Zeeman energy should be greater under the experimental conditions of Du \emph{et al}.'s setup compared to the conditions of Kukushkin \emph{et al}.'s setup, with the exception of the $^2 CF_{(-4)}$-to-$^2 CF_{(-3,-1)}$ transition (see Fig.~\ref{modifiedtheoryvsexperiment}). Nevertheless, the finite thickness correction clearly does not account for the discrepancy between the theoretical predictions and the experimental measurements. Other factors which have been neglected are LL mixing, finite temperature effects, and the effect of sample impurities and disorder. Another source of error is that the precise value for the Coulomb energy in the thermodynamic limit depends on how one does the extrapolation: linear and quadratic regression, or regression taking into account the relative errors on each data point, or not. This issue is particularly acute for the $n=4$ case because, according to Eq.~(\ref{criticalzeeman}), we must multiply the differences in Coulomb energy by a factor of 4 to obtain the prediction for the critical Zeeman energy.

Turning to the results for the analogous states occurring in the 2nd LL, we can deduce an interesting prediction: it would appear that for $n=4$ and $n=3$ the fully polarized states ($^2 CF_{-4}$ and $^2 CF_{-3}$, respectively) lie lower in energy that any of the non spin-polarized states with the same filling factor, although for the $n=2$ case the results are inconclusive. For the states at filling factor $2+3/5$ and $2+4/7$ and perhaps for the states at filling factor $2+2/3$ as well, the prediction of CF theory, therefore, would be that it would not be energetically favorable to depolarize the spin as the field strength is reduced from its highest value and we would expect to observe no spin transitions at all in these cases. An alternative possibility, which is particularly likely for the $2+2/3$ case, is that the energy differences are very small, and that would imply a very low value of the critical field as compared with the analogous cases in the LLL. An experimental observation of extensive non-polarized behavior at filling factors  $2+2/3,2+3/5$ or $2+4/7$ would suggest the existence of ground-state wavefunctions not predicted by CF theory.

The most recent experimental studies of the spin polarization of FQHE states have focused on the question of the spin polarization of the $\nu=5/2$ state.\cite{stern2012,tiemann2012} From the perspective of CF theory, the 5/2 sate is described by CFs with zero effective magnetic field \cite{jainbook}. Given our above conclusion that 2nd LL states described by CF theory would tend to favor a spin polarized configuration, we speculate that a spin polarized configuration may also be favorable at $5/2$.  This would be in agreement with exact diagonalizations of smaller systems\cite{morf1998,rezayi2011}.

\emph{Note added}. As this manuscript was being prepared for publication we became aware of some very recent experiments to determine the spin polarization of quantum Hall states in the 2nd LL at filling factor $2+2/3$, (see Ref. \onlinecite{pan2012}). In that work, evidence is presented indicating that a spin transition of the form we have been discussing here does occur at filling $2+2/3$. The result seems to agree with the predictions of CF theory that we have enumerated here; however, it should also be pointed out that the ground-state wavefunctions may not necessarily be of the CF type (see Ref. \onlinecite{morf1995}).

\acknowledgements

This research was supported by EPSRC Grants No. EP/I032487/1 and No. EP/I031014/1.



\appendix

\section{Geometry}
\label{geometry}

The FQHE occurs in 2D systems in a perpendicular magnetic field. Real systems are, of course, finite in size and, thus will have edges; however, for our investigation, we are concerned only with the bulk properties of the ground-state and do not want to take into account edge effects. One method to eliminate edge effects, from a theoretical perspective, \cite{haldane1983} is to place the quantum Hall system on the surface of a hypothetical sphere of radius $R_S$. The surface of a sphere is described by a spherical co-ordinate system $\Omega \equiv (\theta,\phi)$ with a fixed radius $R_S$. Out of convenience we choose to write wavefunctions in this geometry using a pair of complex spinor co-ordinates $u,v$ such that
\[
u=\cos \left(\tfrac{\theta}{2}\right) e^{i\phi/2} \,\,\, , \,\,\, v=\sin \left(\tfrac{\theta}{2}\right) e^{-i\phi/2}.
\]

A magnetic field $B$ perpendicular to the surface of the sphere can be realized by placing a magnetic monopole at the center of the sphere: the total magnetic flux is $N_\Phi \phi_0= 4 \pi R^2_S B$ (here $\phi_0$ is the flux quantum $\phi_0 = hc/e$). The radius of the sphere is then $R_S=\sqrt{N_\Phi /2}$, in units of the magnetic length
\[
l_0 = \left( \frac{\hbar c}{eB} \right)^{1/2}.
\]
The flux is related to the filling factor by $N_\Phi = N/\nu -S$, where $N$ is the number of electrons and $S$ is called the shift. For spin-polarized CF states we have already derived (in Sec. \ref{wavefunctions}) the result $N_\Phi = 2p(N-1) + 2Q$, where $Q=\pm \tfrac{N-n^2}{2n}$. For the non-spin polarized CF states we can use Eq.~(\ref{effectiveflux}), along with the definition $N=N_{\uparrow}+N_{\downarrow}$, to find an expression for $Q$ in terms of $N$,
\begin{equation}
\label{expressionforq}
Q= \pm \frac{N-n^2_{\uparrow} - n^2_{\downarrow}}{2(n_{\uparrow} + n_{\downarrow})}.
\end{equation}
The total flux is then again given by $N_\Phi = 2p(N-1) + 2Q$.

In each case, in the thermodynamic limit, the sphere radius tends to infinity and so we recover a 2D plane geometry but without edge effects. The length scale set by the magnetic length $l_0$ is generally much smaller than the sphere radius, and so we argue that the physics remains independent of the system size and, therefore, taking the thermodynamic limit is valid.



\section{Degree of Spin Polarization and Proportion of Flipped Spins}
\label{extrastuff}

In this appendix we shall derive two results used in the paper: the degree of spin polarization $\gamma$ and the proportion of spins $\tau$ which must flip when a spin transition takes place.

\subsection{Degree of Spin Polarization}
The degree of spin polarization is defined by
\[
\gamma=\frac{N_{\uparrow}-N_{\downarrow}}{N_{\downarrow}+N_{\uparrow}}.
\]
 Using Eq.~(\ref{effectiveflux}) and the fact that $N=N_{\uparrow}+N_{\downarrow}$ we can deduce for example that
\[
N_{\uparrow}- N_{\downarrow}=N - \frac{2n_{\downarrow}}{n_{\downarrow}+n_{\uparrow}}  (  N + n_{\uparrow} (n_{\downarrow} -n_{\uparrow} ) ),
\]
and so we have
\[
\gamma=\frac{n_{\uparrow}-n_{\downarrow}}{n_{\downarrow}+n_{\uparrow}}\left (1+\frac{2n_{\downarrow}n_{\uparrow}}{N} \right ).
\]
In the thermodynamic limit ($N \rightarrow \infty$) we recover Eq.~\ref{degreeofspin}.

\subsection{Proportion of Flipped Spins}

The proportion of electrons, $\tau$, occupying the effective LL labelled by index $n'$ and with $2Q$ effective flux, is given by
\[
\tau = \frac{2 (|Q|+n')+1 }{N}.
\]
Substituting the expression for $Q$ given in Eq.~(\ref{expressionforq}) we have
\[
\tau = \frac{2}{N} \left ( \frac{N-n^2_{\uparrow} - n^2_{\downarrow}}{2(n_{\uparrow} + n_{\downarrow})} +n' +\frac{1}{2}\right).
\]
In the thermodynamic limit  ($N \rightarrow \infty$) this expression reduces to
\[
\tau =\frac{1}{n_{\uparrow} + n_{\downarrow}}=\frac{1}{n},
\]
which is independent of $n'$.



\section{Monte Carlo Algorithm for a Quantum Hall Fluid}
\label{montecarlo}

In this work we apply the Metropolis Monte Carlo algorithm \cite{newmanbook} to evaluate the ground-state energy of various trial wavefunctions. We wish to evaluate expectation values of operators $\hat{A}$ with respect to the co-ordinate wavefunctions $\psi \left( {\mathbf r}_1 , ... {\mathbf r}_N\right)$:
\[
\left\langle \hat{A} \right\rangle = \frac{\int d {\mathbf r}_1 ...d {\mathbf r}_N \psi^{*} \left( {\mathbf r}_1 , ... {\mathbf r}_N\right) \hat{A} \psi \left( {\mathbf r}_1 , ... {\mathbf r}_N\right)}{\int d {\mathbf r}_1 ...d {\mathbf r}_N \left|\psi \left( {\mathbf r}_1 , ... {\mathbf r}_N\right) \right|^2}.
\]
The Metropolis Monte Carlo procedure works by statistically sampling configurations of the co-ordinates $\left\{ {\mathbf r}_1 , ... {\mathbf r}_N\right\} $ drawn from the probability distribution
\[
\rho_N \left( {\mathbf r}_1 , ... {\mathbf r}_N\right)= \frac{\left|\psi \left( {\mathbf r}_1 , ... {\mathbf r}_N\right) \right|^2}{\int d {\mathbf r}_1 ...d {\mathbf r}_N \left|\psi \left( {\mathbf r}_1 , ... {\mathbf r}_N\right) \right|^2}.
\]
The expectation value of the operator is then estimated using $N_{\mbox{\small s}}$ sets of co-ordinate samples:
\[
\left\langle \hat{A} \right\rangle = \frac{1}{N_{\mbox{\small s}}} \sum_{i=1}^{N_{\mbox{\tiny s}}} \psi \left( {\mathbf r}_1 , ... {\mathbf r}_N\right)^{*} \hat{A} \psi \left( {\mathbf r}_1 , ... {\mathbf r}_N\right).
\]
It can be shown that the standard deviation behaves as $\sigma/<\hat{A}>  \sim 1/ \sqrt{N_{\mbox{\small s}}}$. In our simulations we typically used $N_{\mbox{\small s}} \sim \mathcal{O}(10^7) $.



\section{Algorithm for Numerical Evaluation of Composite Fermion States in Negative Effective Field}
\label{CFalgorithm}

Composite fermion wavefunctions have been intensely scrutinized using MC methods. \cite{jain1997,park2001,moller2005} The principle difficulty for CF wavefunctions is the procedure for LLL projection. The method for doing this projection for CF states in negative effective field was introduced in Ref.~\onlinecite{moller2005}, and the resulting form of the generalized spherical harmonics is repeated here in Eq.~(\ref{CFfinal}). Once written in this form, the key difficulty lies in the evaluation of the multiple derivatives of the Jastrow factor,
\[
\left[ \left(\frac{\partial}{\partial u_i} \right)^{|Q|+m+s} \left(\frac{\partial}{\partial v_i}\right)^{|Q|-m+n-s} J^p_i\right],
\]
with
\[
J_i = \prod_{j \ne i} \left( u_i v_j - u_j v_i \right).
\]
A procedure for evaluating these derivatives is given in Ref.~\onlinecite{parkphd} (see also Ref.~\onlinecite{jainbook}), and this method is also used in Ref.~\onlinecite{moller2005}. Briefly, the method is as follows: first, we pull the Jastrow factor through the derivatives and write
\[
J^p_i \left[ \hat{U}^{|Q|+m+s}_i \hat{V}_i^{|Q|-m+n-s} 1\right],
\]
where
\[
 \hat{U}_i = J^{-p}_i \frac{\partial}{\partial u_i} J^p_i \,\,\, , \,\,\, \hat{V}_i = J^{-p}_i \frac{\partial}{\partial v_i} J^p_i ;
\]
we then introduce
\[
f_i (\alpha,\beta) = \sum_{k=1}^N \left( \frac{v_k}{u_i v_k - v_i u_k} \right)^{\alpha} \left( \frac{-u_k}{u_i v_k - v_i u_k} \right)^{\beta},
\]
from which one can deduce the recursion relations
\[
\begin{array}{c}
\frac{\partial}{\partial u_i} f_i (\alpha,\beta) = - (\alpha +\beta) f_i (\alpha +1,\beta), \\
\frac{\partial}{\partial v_i} f_i(\alpha,\beta) = - (\alpha +\beta) f_i (\alpha,\beta+1). \\
\end{array}
\]
Using these results, one can calculate a series of relations, for example,
\[
\hat{U}_i 1 = p f_i(1,0) \,\,\, , \hat{U}^2_i 1 = p^2 f_i(1,0)^2 - p f_i(2,0) \,\,\, , \ldots.
\]

For CF wavefunctions in negative effective field we must take up to $2|Q|+n = \left( \tfrac{N}{n} \right) -2n$ derivatives with respect to both $u_j$ and $v_j$. As we take more derivatives, the results of this method become increasingly complicated, particularly for smaller values of $n$.

We have determined an alternative method to evaluate the derivatives, which is at its most effective in precisely the regime where the current method runs into difficulties. We shall present our result for the case of $p=1$ only; however, higher $p$ cases could be constructed along the same lines.

Let us re-state exactly what we need to evaluate,  leaving out the un-important constant factors as they can be absorbed into the normalization---for each element of a N-by-N Slater matrix we need to evaluate
\begin{widetext}
\begin{equation}
\hat{Y}^{Q}_{n',m}(u_i,v_i) J_i \propto  \sum_{s=0}^{n'} (-1)^{s}\left( {\begin{array}{*{20}{c}}{n'}  \\s  \\
\end{array}} \right)\left( {\begin{array}{*{20}{c}}
   {2\left| Q \right| + n'}  \\
   {\left| Q \right| + m + s}  \\
\end{array}} \right) u^{s}_i v^{n'-s}_i \left( \frac{\partial}{\partial u_i}  \right)^{|Q|+m+s} \left( \frac{\partial}{\partial v_i} \right)^{|Q|-m+n'-s} \prod_{j \ne i} \left( u_i v_j - u_j v_i \right) .
\end{equation}
It is insightful to expand out the $J_i$ product into a sum as follows:
\[
J_i = v_i^{N-1} \prod_{j \ne i}^{N} u_j - u_i v_i^{N-2} \sum_{j \ne i}^N \left(v_j\prod_{k \ne i,j} u_k \right) +... = \left( \prod_{j \ne i}^{N} u_j \right)\left( \sum_{t=0}^{N-1} (-1)^t e^i_t  v_i^{N-1-t} u_i^{t} \right),
\]
where $e^i_t$ denotes the degree $t$ elementary symmetric polynomial in the $N-1$ variables $v_{j} / u_{j}$ for $j \ne i$,
\[
e^i_t \equiv e_{t,N-1} (v_1 /u_1, ...v_{j} /u_{j}..., v_N / u_N),
\]
for $j \ne i$. The elementary symmetric polynomials are defined by
\begin{eqnarray}
\label{esp}
& & e_{m,N} \left( {x_1 ,...,x_N } \right) =  \begin{cases}
 \sum\limits_{0 < i_1  < i_2  < ... < i_m  \le N} {x_{i_1 } ...x_{i_m } } & m \le N \\
\,\,\,\,\,\,\,\,\,\,\,\,\,\,\,\,\,\,0 & {\rm{otherwise}}. \\
 \end{cases} \nonumber
\end{eqnarray}
With this form for $J_i$ we now evaluate all of the necessary derivatives, which leaves us with:
\begin{equation}
\begin{array}{l}
\hat{Y}^{Q}_{n',m}(u_i,v_i) J_i \propto  \sum_{s=0}^{n'} (-1)^{s}\left( {\begin{array}{*{20}{c}}{n'}  \\s  \\
\end{array}} \right)\left( {\begin{array}{*{20}{c}}
   {2\left| Q \right| + n'}  \\
   {\left| Q \right| + m + s}  \\
\end{array}} \right) \times \\ \\ \sum_{t=|Q| +m +s}^{N-1-(|Q| - m + n' -s)} e^i_t (-1)^t \frac{(N-1-t)!}{N-1-t-(|Q|-m+n'-s)!} v_i^{N-1-t-(|Q|-m)} \frac{t!}{(t-(|Q|+m+s))!} u_i^{t-(|Q|+m)}.
\end{array}
\label{newform}
\end{equation}

The expression in Eq.~(\ref{newform}) appears complicated at first sight, but notice that the sum over $t$ contains at most $N-1 -\left( \tfrac{N}{n} \right) +2n$ terms. If the CF state only fills a small number of effective LLs (i.e., $n=1,2,...$), then the number of terms in the sum is actually quite small.

A key result is that the elementary symmetric polynomials can be calculated recursively using one of Newton's identities, \cite{macdonaldbook}
\[
e_{m,N} \left( {x_1 ,...,x_N } \right) = \frac{1}{m} \sum_{r=1}^{m} (-1)^r p_{r,N} \left( {x_1 ,...,x_N } \right) e_{m-r,N} \left( {x_1 ,...,x_N } \right) ,
\]
where $p_{r,N} \left( {x_1 ,...,x_N } \right) = \sum_{i=1}^N x_i^r$ are the power-sum polynomials. Also, note the following recursive identity:
\[
e_{m,N-1} \left( {x_1 ,...,x_{j \ne i},...,x_N } \right) = e_{m,N} \left( {x_1 ,...,x_N } \right) - x_i e_{m-1,N-1} \left( {x_1 ,...,x_{j \ne i},...,x_N } \right).
\]
\end{widetext}
Wielding these two identities it is possible to build an efficient algorithm to generate the full set of $e^i_t $ for $i=1,...,N$ that is required to calculate the elements of the form $\hat{Y}^{Q}_{n',m}(u_i,v_i) J_i $ for the $N\times N$ Slater matrix.

Once the Slater matrix is populated, the remainder of the work is involved in calculating its determinant. The full algorithm evaluates the probability density of a given CF state for a given set of co-ordinates. Based on the times recorded when running the program, our full algorithm evaluation time scales roughly as $N^{2.2}$--$N^{2.75}$, the better figure occurring for $n=1$ and the worse figure for $n=4$, which confirms our assertion that our algorithm is more efficient for smaller values of $n$.

We note that the evaluation of the CF wavefunctions tends to suffer from a numerical precision issue, particularly for large $N$. To overcome the issue we simply store all numerical values to a higher precision---although this slows down the algorithm considerably, we are able to obtain accurate results up to around $N=40$, which is high enough for our purposes.




\begin{thebibliography}{99}
\bibitem{tilt}  In principle, tilting the magnetic field can alter the effective interaction between the electrons by modifying the form of the wavefunction in the direction perpendicular to the electron layer.    We have not accounted for this effect.  However, we note that the in-plane magnetic length is never much less than the well width so this effect is not expected to be substantial  (see Table \ref{valuesofd}).
\end{thebibliography}
\end{document}